\documentclass[aps,prd, nofootinbib,floatfix,amssymb]{revtex4} 
\usepackage{epsf}
\usepackage{graphicx,epsfig}
\usepackage{amsfonts}
\usepackage{amssymb}
\usepackage{cancel}

\def\gsim{\mathrel {\vcenter {\baselineskip 0pt \kern 0pt \hbox{$>$} \kern 0pt \hbox{$\sim$} }}}

\def\ba{\begin{eqnarray}}
\def\ea{\end{eqnarray}}

\def\lsim{\mathrel {\vcenter {\baselineskip 0pt \kern 0pt \hbox{$<$} \kern 0pt \hbox{$\sim$} }}}

\def\gsim{\mathrel {\vcenter {\baselineskip 0pt \kern 0pt \hbox{$>$} \kern 0pt \hbox{$\sim$} }}}

\newcommand{\dho}{\hat{{\bf \Omega}}}

\begin{document}

\title{3-pt Statistics of Cosmological Stochastic  Gravitational Waves}
\author{Peter Adshead}
\affiliation{Department of Physics\\ Yale University, New Haven, CT 06511 USA}
\author{Eugene A. Lim}
\affiliation{Department of Physics and ISCAP,  \\ Columbia University, New York, NY 10027 USA}
\preprint{CAS-KITPC/ITP-156}

 \begin{abstract}
 
We consider the 3-pt function (i.e. the bispectrum or non-Gaussianity) for stochastic backgrounds of gravitational waves. We estimate the amplitude of this signal for the primordial inflationary background, gravitational waves generated during preheating, and for gravitational waves produced by self-ordering scalar fields following a global phase transition.  To assess detectability, we describe how to extract the  3-pt signal from an idealized interferometric experiment and compute the signal to noise ratio as a function of integration time. The 3-pt signal for the stochastic gravitational wave background generated by inflation is unsurprisingly tiny.  For gravitational radiation generated by purely \emph{causal}, classical mechanisms we find that, no matter how non-linear the process is, the 3-pt correlations produced vanish in direct detection experiments. On the other hand, we show that in scenarios where the B-mode of the CMB is sourced by gravitational waves generated by a global phase transition, a strong 3-pt signal among the polarization modes could also be produced. This may provide another method of distinguishing inflationary B-modes. To carry out this computation, we have developed a diagrammatic approach to the calculation of stochastic gravitational waves sourced by scalar fluids, which has applications beyond the present scenario.

\end{abstract}
 \maketitle

\section{Introduction}

Great strides are being made in both the technological and theoretical aspects of the direct detection of astrophysical gravitational waves. Ground based interferometers like LIGO \cite{LIGO} have achieved their design sensitivities and are in the process of being upgraded to even greater precision \cite{Smith:2009bx}. Design of a  space-based detector, LISA \cite{LISA}, is well underway with a pathfinder mission scheduled for launch in mid-2011 and a possible launch date for the full mission in the next decade or so.  Further, planning for the next generation space-based gravitational wave detectors, BBO and DECIGO, has begun \cite{Sato:2009zzb, Harry:2006fi}. These detectors are specifically designed to search for \emph{cosmological stochastic gravitational waves} (SGW). Beyond direct detection experiments,  large scale B-mode polarization of the cosmic microwave background   \cite{Kamionkowski:1996ks} can be sourced by a spectrum of long wavelength gravitational waves at last scattering.  The Planck mission will constrain a tensor to scalar ratio on the order of $r\sim 0.1$ while work is well underway on proposals \cite{Baumann:2008aq} that will probe to $r \sim 0.01$.

A stochastic background of gravitational waves can be generated in a variety of ways. Unresolved point sources such as neutron star or black hole binary systems generate a stochastic background in the confusion limit \cite{ Schneider:2000sg}. Quantum fluctuations of the metric during inflation \cite{ Mukhanov:1981xt, Guth:1982ec, Abbott:1984fp, Starobinsky:1982ee} are amplified on super Hubble scales, generating a stochastic background whose amplitude directly probes the energy scale of inflation. Following inflation, explosive particle production associated with a phase of pre/reheating would also produce a stochastic background  \cite{Khlebnikov:1997di, Easther:2006gt, Easther:2006vd, Easther:2007vj, Dufaux:2008dn,Dufaux:2007pt,GarciaBellido:2007dg, GarciaBellido:2007af, Price:2008hq, Easther:2008sx}.  Further, SGW can be generated by phase transitions \cite{Kosowsky:1992rz, Kamionkowski:1993fg, Kosowsky:2001xp, JonesSmith:2007ne, Fenu:2009qf, Caprini:2009yp}, bubble collisions \cite{Kosowsky:1991ua, Kosowsky:1992vn} or even more exotic processes involving warped extra dimensions \cite{Randall:2006py}. 

Recently, models have been proposed where a \emph{scale-invariant} spectrum of 
SGW,  mimicking the inflationary spectrum on subhorizon scales, is generated by the relaxation of a disordered scalar matter\footnote{In this paper we will deal exclusively with scalar matter although this is not a necessary condition.} field with a white noise spectrum on superhorizon scales. Such a white noise spectrum can be laid down by a global phase transition event, for example \cite{Krauss:1991qu,JonesSmith:2007ne,Fenu:2009qf}. More generally, non-equilibrium processes in cosmology produce scale-dependent spectra of gravitational waves. This should not be a surprise: violent motion of large masses generates a non-trivial quadrupole moment.

In the study of cosmological density perturbations in large scale structure and in the Cosmic Microwave Background, such higher correlations of perturbations are called non-Gaussianities \cite{Bartolo:2004if} -- any Gaussian field is completely described by its power spectrum, or 2-pt function.\footnote{For a completely Gaussian field,  all correlation functions of an odd number of fields vanish and all even correlations can be written as products of the power spectrum. Such truncated correlation functions are usually called \emph{disconnected}, see for example \cite{Malaspinas:2004rq}.} 
Indeed, while inflation is expected to produce a highly Gaussian spectrum of SGW akin to that of the spectrum of density perturbations \cite{Komatsu:2008hk}, cosmological SGW foregrounds from active sources are expected to be highly non-Gaussian. However, these processes occur on characteristic scales which are smaller than the size of the horizon. We present a simple argument that any \emph{causal} process, i.e. one which operates inside a single post-inflationary horizon volume, which generates gravitational waves predicts a vanishing 3-pt correlation in our detectors. Such processes include preheating (which we will discuss in detail as an example below), bubble collisions, and all of the other processes mentioned above. Inflation on the other hand predicts correlations among all modes at all scales.

In addition to the present work, the study of non-Gaussian features in SGW spectra has received little attention: Drasco and Flanagan investigated the statistics of popcorn noise \cite{Drasco:2002yd}, while Seto has suggested the use of the 4-pt correlator to study intermittent bursts \cite{Seto:2008xr, Seto:2009ju}. Racine and Cutler \cite{Racine:2007gv} investigated deviations from Gaussianity of unresolved galactic white dwarfs binaries and suggested that it is small due to the large number of sources. 

This paper is organized as follows.  In Section \ref{sect:observations} we discuss sources of 3-pt correlation functions and their prospects for detection -- in Section \ref{sect:obsinf} we consider the 3-pt signal generated by quantum effects during inflation while in Section \ref{sect:obspre} we estimate the expected amplitude of the 3-pt function from active scalar sources. We consider two scenarios, global phase transitions and preheating, and present some preliminary analytical results.  The details of the calculations are left to the Appendices. In Appendix \ref{sect:3pt} we construct the 3-pt estimator of SGW, given a set of detector data streams. In Appendix \ref{app:scalarsources} present a diagrammatic method of calculating general $N$-pt SGW correlation functions from linear scalar sources which will have applications beyond the present scenario.  Finally, we conclude in Section \ref{sect:conclusions}.

\section{Sources and Observations} \label{sect:observations}

Cosmological SGW can be divided into those sourced by initial quantum fluctuations (namely those laid down during inflation) and those sourced classically by non-zero quadrupole moments. Gravitational waves from inflation, at the moment of creation, behave like free fields with Gaussian initial conditions and hence possess a vanishing 3-pt correlation. Non-zero 3-pt correlations are formed when these fields interact gravitationally (see the left graph in Fig. \ref{fig:GW3pt}) and carry information about the gravitational coupling strength $H/M_p$. Inflationary gravitational waves are correlated on all scales due to their creation during an epoch of accelerated expansion. These correlations are laid down as the scales leave the horizon and are frozen until reheating. On the other hand, actively sourced gravitational waves are generated by gravitational \emph{bremsstrahlung}, and carry information about the quadrupole moment of the source  (see the right graph in Fig. \ref{fig:GW3pt}). These gravitational waves are laid down as each scale enters the horizon, and are only correlated on scales comparable to the horizon size at the time they were created. We consider two specific models, preheating and the global phase transition scenario of Jones-Smith et al. \cite{Krauss:1991qu,JonesSmith:2007ne}. 

A 3-pt correlation of gravitational waves is produced during inflation via the right graph in Fig. \ref{fig:GW3pt}. However, this is highly suppressed as each internal scalar line is a copy of the scalar power spectrum, supressing the graph by a factor of $\mathcal{O}(\mathcal{P}(k)^{3}) \sim 10^{-30}$. 

\subsection{Inflation}\label{sect:obsinf}

\begin{figure}
\centering
\includegraphics[scale = 0.6]{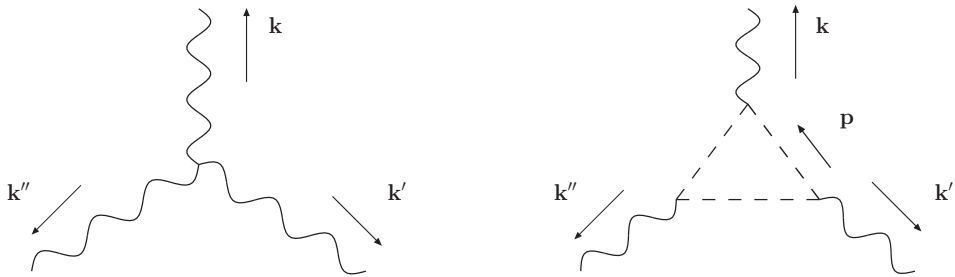}
\caption{Non-trivial 3-pt correlations of gravitational waves can be generated directly from graviton-graviton interactions (left graph) or indirectly via interactions with scalars (right graph). Both processes are always present, however, direct interaction dominates the 3-pt correlation during inflation while for scalar sourced gravitational waves this process is highly suppressed relative to the loop.  \label{fig:GW3pt}}
\end{figure}

Even if inflation is exactly de Sitter, the spectrum of SGW it generates will be non-Gaussian, due to gravitational self-interaction. At leading order, this non-Gaussianity is sourced by a 3-pt interaction term, represented diagrammatically in Fig. \ref{fig:GW3pt} (left graph). The amplitude of this process is calculated using the ``in-in'' formalism\footnote{In general, these amplitudes will also be corrected by both scalar and GW loops, but they will be subdominant -- see for example \cite{Adshead:2009cb}.} \cite{Maldacena:2002vr}. 

Each external GW leg is ${\cal O}(H/M_p)$ due to canonical normalization of the graviton, while the coupling term is purely gravitational, and also ${\cal O}(H/M_p)$. Thus, after freeze out, gravitational waves outside the horizon will possess a non-zero 3-pt correlation function, with amplitude $(H/M_p)^4$. 

In \emph{any} model of inflation in purely Einstein gravity, the 3-pt correlation function of gravitational waves at freeze-out is \cite{Maldacena:2002vr}
\begin{eqnarray}
\langle h^A({\bf k}) h^{A'}({\bf k}')h^{A''}({\bf k}'')\rangle_{\rm inf} &=& \left(-K + \frac{kk'+k'k''+kk''}{K}+\frac{kk'k''}{K^{2}}\right) (2\pi)^3\delta ({\bf k}+{\bf k}'+{\bf k}'')  \nonumber \\
&& \times \frac{H_{*}^4}{M_{\rm pl}^4}\frac{-4}{8(kk'k'')^3}(e_{ii'}^{A}({\bf k})e_{jj'}^{A'}({\bf k}')e_{ll'}^{A''}({\bf k}'')t_{ijl}t_{i'j'l'}), \label{eqn:maldacena}
\end{eqnarray}
where $H_{*}$ is the Hubble rate at horizon crossing, $K = k+k'+k''$, and the tensorial structure is $t_{ijk} \equiv k'{}^{i}\delta_{jl}+k''{}^{j}\delta_{il}+k^l\delta_{ij}$. In this work, since the exact details of the polarization sum in Eqn. (\ref{eqn:maldacena}) do not concern us, we replace it with the appropriate powers of $k$ from dimensional analysis.

To calculate the present amplitude of the inflationary 3-pt correlation (Eqn. (\ref{eqn:maldacena})) we evolve the metric perturbations from horizon re-entry using the gravitational wave transfer function.\footnote{In reality the GW transfer function is much more complicated \cite{Boyle:2005se}, however, the corrections are $\mathcal{O}(1)$ and so for this work we neglect them. Also, one might worry that the different reentry time for each mode will induce a phase shift in the 3-pt correlation today. However, it is easy to show that for modes reentering in the radiation era, the phase shift in the signal 3-pt is an overall constant and hence can be set to zero.} For modes reentering the horizon during radiation domination, $\Delta_{gw} = z_{eq}^{-1/2}(H_0/2\pi f)$, replacing ${\bf k} =2\pi f \hat{{\bf \Omega}}$. We find
\begin{equation}
{\cal F}(f,f,f'') _{\rm today} ={\cal F}(f,f,f'') _{\rm inf} \times \left(\frac{H_0}{2\pi}\right)^3\frac{1}{ff'f''} \frac{1}{z_{eq}^{3/2}}, \label{eqn:transfer}
\end{equation}
where  ${\cal F}(f, f', f'')$ is defined by 
\begin{eqnarray}
\langle h(f, \dho)h(f', \dho')h(f'', \dho'')\rangle = \mathcal{F}(f,f',f'')\delta(f\dho+f'\dho'+f''\dho'').
\end{eqnarray}
Here, $z_{eq} \approx 4\times 10^3$ is the redshift of matter-radiation equality and $H_0$ is the Hubble constant today. As detailed in Appendix \ref{sect:3pt}, the total integration time $T$ of a direct detection experiment can be chopped into ``chunks'' of identical duration $\Delta T \sim f_*$, where $f_*$ is roughly the frequency of minimum noise. We can then construct the 3-pt estimator per chunk and, since the noise is uncorrelated across chunks, the total signal to noise scales as $\sqrt{M}$, where $M=T/\Delta T$ is the total number of chunks:
\begin{equation}
\mathrm{SNR} =\mathrm{shape}\times \left(\frac{H_{\rm inf}}{M_{\rm pl}}\right)^{4} \left(\frac{H_0}{2\pi}\right)^3\left(\frac{1}{f_*}\frac{1}{z_{eq}^{1/2}}\right)^3\left(\frac{1}{N^{2}_1(f_*)N^{2}_2(f_*)N^{2}_3(f_*) (\Delta f)^3}\right)^{1/2}  \times \sqrt{M},
\end{equation}
where $\Delta f$ is the width of the window in frequency space. Here we collect the frequency integral into a dimensionless quantity we call ``shape''\footnote{This nomenclature follows standard convention in the studies of CMB non-Gaussianities, where the the 3-pt is similarly integrated over all possible triangles to produce a single measure, see for example \cite{Babich:2004gb}.}
\begin{equation}
\mathrm{shape} \equiv
\int^{\theta_{\wedge}^{max}}_{\theta_{\wedge}^{min}} d\cos \theta_{\wedge} \int_{f_*-\Delta f}^{f_*+\Delta f} f^{2}df f'^{2}df' f''^{2}df'' \left(\frac{H_{*}}{M_{\rm pl}}\right)^{-4}{\cal F}(f,f,f'') _{\rm inf}\frac{8\pi^2}{f^2}\delta(f-\sqrt{f'{}^2+f''{}^2+2f'f''\cos \theta_{\wedge}}),
\end{equation}
where $\theta_{\wedge}$ is the angle between $\dho'$ and $\dho''$. The shape encodes where the 3-pt signal has support in the 3 dimensional parameter space $(f,f',f'')$ -- different gravitational wave generation mechanisms will, in general, produce different shape functions. The shape can be  integrated numerically, but we note that the filters pick up triangles that are roughly equilateral, and approximate the shape simply as
\begin{eqnarray} \label{eqn:shape}
\mathrm{shape} & \equiv & \frac{1088}{9}  \pi^{2}\frac{\left(f_*^2+\Delta
   f^2\right)}{\left(f_*^2-\Delta f^2\right)^2} \frac{\Delta f^{3}}{f_{*}}.
\end{eqnarray}

Given a total integration time $T$, we obtain $M =  T/(10 \times f_*^{-1})$ independent observations, including a fudge factor of $10$ to reduce edge effects from chopping the signal. The total integration time expected for a $90\%$ confidence detection then scales like (assuming identical, time independent and constant detector noise)
\begin{equation}
T \propto f_*^{11}N^6(f_*)(\Delta f)^{-6}.
\end{equation}
In other words, as in the detection of the SGW power spectrum, \emph{low noise at a low target frequency $f_*$ is advantageous.}

An instrument like BBO is  designed to detect the inflationary power spectrum, but it is highly unlikely that we will be able to detect the 3-pt correlation -- this is a consequence of the fact that inflation is  highly \emph{Gaussian}. For a typical BBO/DECIGO detector with $N(f_*) \sim 10^{-24}\, \mathrm{Hz}^{-1/2}$ at $f_* \sim 0.1$ Hz and $\Delta f \sim 0.033$ Hz, and assuming GUT scale inflation $H_{\rm inf}/M_{\rm pl}\sim 10^{-6}$, $T \sim 0.27\times 10^{6}h^{-6}$ years where $ H_0 = 100h\; \mathrm{km/s}/\mathrm{Mpc}$. It may appear that, since $T$ scales with $f_*^5N(f_*)^6$ (assuming $f_* \sim \Delta f$), if we were to increase the sensitivity of BBO by one order of magnitude or move $f_*$ down a decade in frequency, we would be able to detect the 3-pt correlation on a reasonable timescale. However, this result scales with $(H_{\rm inf}/M_{\rm pl})^{-8}$.  Unless the energy scale of inflation is near the upper limit currently allowed by WMAP \cite{Komatsu:2008hk} measurements of the cosmic microwave background, the inflationary 3-pt function will be well out of reach of direct detection experiments.

\subsection{Active Scalar Sources}\label{sect:obspre}

Beside the stochastic gravitational waves generated by the amplification of quantum fluctuations during inflation, gravitational waves can be generated \emph{classically} via violent motion of mass-energy. For example, a period of turbulent cosmological evolution, say during preheating, will result in the generation of large gradients and the copious production of SGW.

``Active sources'' refers to sources which are physically moving.  We focus on the results here and present a general review of scalar sourced gravitational waves in Appendix \ref{app:scalarsources}. We will see that for all causally generated, actively sourced, gravitational waves, the 3-pt function vanishes in direct detection experiments. Gravitational waves from preheating will not be detectable in the CMB, since the long wavelength modes necessary to generate the temperature and polarization anisotropies are not produced, but global phase transition models continuously sources horizon scale gravitational waves. These modes will affect the CMB anisotropies on scales smaller than the horizon (in real space) at last scattering. In particular, we will show that the source of the polarization modes in the CMB for this model has large 3-pt correlations (i.e. it is highly non-Gaussian).
 This means that the polarization bispectrum of the CMB, i.e. both the $\langle BBB\rangle$ and the $\langle EBB\rangle$ statistics will be non-trivial for this model, and could provide a powerful test for non-inflationary sources of the polarization modes.

\subsubsection{Preheating and other causal mechanisms}

During inflation, quantum fluctuations of the metric are amplified on super Hubble scales by the accelerated expansion.  Following inflation, in most models, the inflaton decays and reheats the universe. The first stage of this process, preheating, is dominated by an explosive and non-perturbative production of highly inhomogeneous, non-thermal fluctuations of the inflaton and the other fields coupled to it. The inhomogeneous decay of the inflaton and the turbulent phase that follows it are inevitably  accompanied by the production of gravitational waves. This topic was first discussed by Khlebnikov and Tkachev \cite{Khlebnikov:1997di} and went largely untouched for almost ten years.\footnote{See also \cite{Bassett:1997ke}.} Recently a flurry of papers has appeared with more accurate numerical simulations improving and expanding on earlier work \cite{Easther:2006vd, Easther:2006gt, Easther:2007vj, Dufaux:2008dn, Dufaux:2007pt, GarciaBellido:2007dg, GarciaBellido:2007af, Price:2008hq}, and exploring the scaling relationship for this signal.

The stochastic gravitational wave spectrum from preheating has a peak amplitude at a physical scale that depends only on the energy density at the end of inflation \cite{Easther:2006gt},
\begin{eqnarray}
l \propto V_{end}^{1/4}.
\end{eqnarray}
This can be expressed as a frequency via
\begin{eqnarray}
f = 6 \times 10^{10}\sqrt{\frac{H_{\rm e}}{M_{\rm pl}}}~\mathrm{Hz},
\end{eqnarray}
where $H_{e}$ is the Hubble rate at the end of inflation. To obtain a peak at $f = 0.1$ Hz requires $H_{e}/M_{\rm pl} \sim 10^{-22}$, or inflation ending near the TeV scale. The peak amplitude of the preheating gravitational waves is (largely) independent of the frequency, and is estimated to be $\Omega_{gw}h^2 \sim 10^{-11}$ today \cite{Easther:2006vd}.

Since preheating is a completely causal process, only modes that are within the horizon are excited. If $k_p$ is the physical scale of preheating then
\begin{equation}
\frac{H_e}{k_p} \geq 1.
\end{equation}
Negligible amounts of gravitational waves are generated outside the horizon, so we do not expect super-horizon correlations. Modes within each Hubble patch are expected to be highly correlated at the 3-pt level, we expect gravitational waves to be \emph{uncorrelated} across patches. Since the signal in our detectors is the sum of  contributions from a large number of uncorrelated patches, a simple application of the central limit theorem implies that their 3-pt correlation function will be highly suppressed relative to the 2-pt function. Specifically, there will be approximately $N \sim (H_e/H_0)^{2} \sim 10^{80}$ patches for TeV scale inflation (and even more for GUT scale inflation). The 3-pt correlation function for preheating must be suppressed by $(H_e/H_0)$, i.e. the square root of the number of different causal disconnected patches in the sky, relative to the 2-pt function.\footnote{This same argument can be applied to compact binary sources: while individually each source is highly non-Gaussian, the central limit theorem tells us that as long as each individual source is uncorrelated the sum will be Gaussian. In principle, although preheating patches ``know'' about the inflaton potential and hence is correlated in some way, the fact that the process is likely to be highly chaotic means that this knowledge is rapidly lost and hence we do not expect the patches to be correlated. Nevertheless, this is not a given -- recently \cite{Bond:2009xx} argued that for some classes of SUSY-inspired models of inflation, large non-Gaussian spikes of the curvature perturbation can be laid down over superhorizon scales which preserve this memory, perturbations which may source a super-horizon spectrum of SGW. We are extremely grateful to Richard Easther and Lam Hui for pointing this out to us.}
We thus expect that SGW backgrounds from preheating would be {\it observed} to be highly Gaussian today. If such a spectrum is detected by direct detection experiments there is the intriguing possibility that this highly gaussian source of SGW may be used as a backlight to probe the foreground structures. We postpone a discussion of this possibility to a future publication.

This argument for Gaussianity does not rely on any of the details of preheating and thus extends to other SGW that are sourced on subhorizon scales by active processes in the early universe.

\subsubsection{Global Phase Transitions and the CMB}\label{sect:globalphasetrans}

It has been suggested that a global phase transition in the early universe can actively source a scale invariant spectrum of stochastic gravitational waves on large scales, mimicking that of inflation \cite{JonesSmith:2007ne}. It has also been suggested that such a process may also mean that the detection of large scale B-mode polarization of the CMB would not be a unique ``smoking gun''  signal of inflation \cite{JonesSmith:2009ga}. Using causality arguments,  Baumann and Zaldarriaga \cite{Baumann:2009mq} have shown the polarization signal from an actively sourced spectrum of gravitational waves would be distinct from that of inflation. The usual B-modes are defined non-locally in terms of the Stokes parameters $Q$ and $U$, and are not obliged to vanish for scales outside the horizon. However, Baumann and Zaldarriaga construct a real space correlation function which respects causality. For inflation, this correlation function has features on scales larger than the size of the causal horizon at last scattering that cannot be present for any mechanism which causally generates gravitational waves during the standard, post-inflationary period of the universe. We will see that, even if a such a scenario can completely mimic inflation at the level of the 2-pt function, they are vastly different at the level of the 3-pt function. This may provide an additional test for determining the inflationary origin of the CMB B-mode polarization signal.

We consider the model of \cite{JonesSmith:2007ne} (see also \cite{Fenu:2009qf}, the formalism of which we adopt for this paper) and present only the key details here. The reader is referred to the original papers for additional information.  A field, $\Phi({\bf x}, t)$, in the vector representation of $O(N)$ is in an initially symmetric state with zero vacuum expectation value (vev), $\langle \Phi({\bf x}, t) \rangle = 0$. The field is maintained in this state by a quadratic potential which possibly arises from thermal corrections, or from a coupling to the inflaton. As the universe evolves, thermal corrections become negligible or inflation ends via a tachyonic instability which causes the potential to evolve into a mexican hat. The field obtains a non-zero vev, $\langle \Phi({\bf x}, t) \rangle  = v$, by ``rolling'' to the true vacuum state. On scales larger than the horizon size, the direction in field space of the vector $\Phi({\bf x}, t)$ is uncorrelated and there is gradient energy associated with the $N-1$ Goldstone modes, $\rho \sim (\partial\Phi)^{2}$. As these modes enter the horizon, the scalar field aligns itself and some of the gradient energy is radiated into gravitational waves.

On large scales (or low energies) the field is confined to the vacuum manifold, $\sum_{a}\phi^{2}_{a}({\bf x}, t) = v^{2}$, and the dynamics of the $N-1$ Goldstone modes are well described by the non-linear sigma model. Furthermore, in the large $N$ limit, the model is  soluble \cite{Turok:1991qq}. Initial conditions (at $\tau = \tau_{*}$) are assumed to be white noise on super horizon scales with vanishing power on subhorizon scales, corresponding to the field being initially aligned on these scales
\begin{eqnarray}
\langle \phi_{a}({\bf k}, \tau_{*})\phi_{b}({\bf k}', \tau_{*})\rangle & = & \left\{\begin{array}{lc}
6\pi^{2}\tau_{*}^{3}(2\pi)^{3}\frac{\delta_{ab}v^{2}}{N}\delta({\bf k}+{\bf k}') &,\; k\eta_{*}\ll 1\\
0 &,\; k\eta_{*} >1.
\end{array}\right.
\end{eqnarray}

It is easy to see why a strong correlation between the 2-pt and 3-pt correlation functions is expected from such a source: both are generated from the same interaction term (Fig \ref{fig:vertex}) at 1-loop. From dimensional analysis,  $M_p^2 h \sim \langle \phi \phi \rangle$, and in Fourier space
\begin{equation}
k^3 \langle h_k^2 \rangle \sim (\langle \phi_k^2 \rangle)^2~,~ k^6 \langle h_k^3 \rangle \sim (\langle \phi_k^2 \rangle)^3.
\end{equation}
Via Wick's theorem for a \emph{linear}  source $\phi$, it follows that 
\begin{equation}
\frac{k^6\langle h_k^3\rangle}{(k^3\langle h_k^2\rangle)^{3/2}} \sim 1.
\end{equation}
In other words, we should expect that the 2-pt and 3-pt (and indeed, any higher point) functions are equally important.\footnote{In the language of the scalar bispecturm, complete correlation between $\langle \zeta \zeta\rangle$ and $\langle \zeta\zeta\zeta\rangle$ would mean $f_{nl}\sim 10^{5}$.}
Since gravitational waves source both $E$ and $B$ modes, we expect that such a mechanism sources non-trivial $\langle BBB\rangle$ and  $\langle EBB\rangle$ correlations on the CMB sky. As both such correlations are vanishingly small in the standard inflationary scenario, measurement of these correlations will be a smoking gun for a non-standard source of polarization\footnote{In fact, this  process is present even during inflation -- gravitons are sourced by bremsstrahlung of the inflaton field itself, but it is easy to show that the contribution to the inflationary 3-pt from this process is highly suppressed.}. We leave the details of the construction of the 3-pt polarization correlation functions to future work.

Let us now return to the detailed calculation, one can skip right ahead to Eqn. (\ref{eqn:GPTanswer}) if one is not interested in the technical details. For power law expansion in conformal time $\tau$, $a\propto \tau^{\beta}$ ($\beta = 1$ for radiation domination, $2$ for matter domination), one finds $\langle \phi_{a}({\bf k}, \tau)\phi_{b}({\bf k}', \tau')\rangle = (2\pi)^{3}\delta_{ab}\delta({\bf k}+{\bf k}')F(k,\tau,\tau')$ where
\begin{eqnarray}\label{eqn:scalarprop}
F(k,\tau,\tau')  & = & 6\pi^2\frac{v^{2}}{N}\frac{\Gamma(\beta+1/2)\Gamma(2\beta+3/2)}{\Gamma(\beta)}(\tau\tau')^{3/2}\frac{J_{1+\beta}(k\tau)}{(k\tau)^{1+\beta}}\frac{J_{1+\beta}(k\tau')}{(k\tau')^{1+\beta}}.
\end{eqnarray}

The 3-pt function induced by this source in the co-located detector approximation is then given by Eqn. (\ref{eqn:gaussianpreheat}), which together with the above ``propagator'' (technically a stochastic average) for the scalar field gives 
\begin{eqnarray}\label{eqn:globalphasetransition} 
&&\langle h_{ij}({\bf k}, f)h_{jk}({\bf k'},f')h_{ki}({\bf k''},f'')\rangle \\\nonumber& = &   -4\mathcal{O}_{ij, mn}({\bf \hat{k}}) \mathcal{O}_{jk, op}({\bf \hat{k}'}) \mathcal{O}_{ki, qr}({\bf \hat{k}}'')\delta({\bf k}+{\bf  k}'+{\bf k}'')\frac{(16\pi G)^{3}}{k k' k''} N\left(6\pi^2\frac{v^{2}}{N}\Gamma(3/2)\Gamma(7/2)\right)^{3}\\ \nonumber&& \times\int_{\tau_{i}}^{\tau_{f}}d\tau_{1} d\tau_{2}d\tau_{3} a(\tau_{1})\tau_{1}^{3}\sin[k\tau_{1}]a(\tau_{2})\tau_{2}^{3}\sin[k'\tau_{2}] a(\tau_{3})\tau_{3}^{3}\sin[k''\tau_{3}]\int\frac{d^{3}p}{(2\pi)^{3}}p_{m}p_{n}\\\nonumber
&&\times \Big(p_{o}p_{p}(p-k)_{q}(p-k)_{r}\frac{J_{2}( |{\bf k}' + {\bf p}|\tau_{3})}{(|{\bf k}' + {\bf p}|\tau_{3})^{2}}\frac{J_{2}(|{\bf k}' + {\bf p}|\tau_{2} )}{(|{\bf k}' + {\bf p}|\tau_{2})^{2}} \frac{J_{2}(p \tau_{1})}{(p\tau_{1})^{2}}\frac{J_{2}(p\tau_{2} )}{(p\tau_{2})^{2}}\frac{J_{2}(|{\bf k} - {\bf p}| \tau_{1})}{(|{\bf k} - {\bf p}|\tau_{1})^2}\frac{J_{2}(|{\bf k} - {\bf p}|\tau_{3} )}{(|{\bf k} - {\bf p}|\tau_{3})^2}\\\nonumber
&& +(p-k)_{o}(p-k)_{p}p_{q}p_{r}\frac{J_{2}( |{\bf k}'' + {\bf p}|\tau_{3})}{(|{\bf k}'' + {\bf p}|\tau_{3})^{2}}\frac{J_{2}(|{\bf k}'' + {\bf p}|\tau_{2} )}{(|{\bf k}'' + {\bf p}|\tau_{2})^{2}}\frac{J_{2}(|{\bf k}-{\bf p}| \tau_{1})}{(|{\bf k}-{\bf p}|\tau_{1})^{2}}\frac{J_{2}(|{\bf k}-{\bf p}|\tau_{2} )}{(|{\bf k}-{\bf p}|\tau_{2})^{2}}\frac{J_{2}(p \tau_{1})}{(p\tau_{1})^{2}}\frac{J_{2}(p\tau_{3})}{(p\tau_{3})^{2}}\Big).
\end{eqnarray}
Here $\mathcal{O}_{ij,kl}({\bf \hat{k}'})$ is the transverse traceless projector, defined in Appendix \ref{app:scalarsources}. An exact analytic evaluation of this expression is difficult. In principle there is no obstacle to numerical integration, but for our purposes it is sufficient to approximate it as follows:
\begin{itemize}
\item{Work in the equilateral limit, $|{\bf k}|=|{\bf k}''|=|{\bf k}''|$, ${\bf \Omega}+{\bf \Omega}'+{\bf \Omega}''= 0$. Most of the power is produced at horizon crossing, so we expect that the signal will be strongly peaked on equilateral shapes.}
\item{Work in the long wavelength limit $k \tau < 1$, $k \tau' < 1$ and $k \tau'' < 1$ for all times between $\tau_{*}$ and $\tau_{end} = 1/k$ so that $\sin(k\tau) \approx k\tau$ etc}
\item{We neglect the angular dependence of $|{\bf k''}+{\bf p}|$ and $|{\bf k}-{\bf p}|$}
\item{We can then use asymptotic expansions of the Bessel functions. In the range $p < \min(1/\tau_{1},1/\tau_{2},1/\tau_{3})$ we use the small argument expansion of the Bessel function. In the region $\min(1/\tau_{1},1/\tau_{2},1/\tau_{3})<p < \max(1/\tau_{1},1/\tau_{2},1/\tau_{3})$ we distinguish between large and small argument expansions and finally in the range $\max(1/\tau_{1},1/\tau_{2},1/\tau_{3})<p<1/\tau_{*}$ we can use the large argument expansion of the Bessel function. }
\end{itemize}
There are two limits in which we can expand the Bessel functions, large and small argument expansions, these are
\begin{eqnarray}
J_{\nu}(x) & \simeq & \left\{\begin{array}{lc}
\frac{x^{\nu}}{2^{\nu}\Gamma(\nu+1)} & \text{for $x \ll 1$},\\
\sqrt{\frac{2}{x\pi}}\cos\left(x - \frac{(2\nu+1)\pi}{4}\right) &  \text{for $x \gg 1$}.
\end{array}\right.
\end{eqnarray}
Under these approximations, we can write
\begin{eqnarray}
&&\langle h_{ij}({\bf k}, f)h_{jk}({\bf k'},f')h_{ki}({\bf k''},f'')\rangle \\\nonumber& \approx &   -4\mathcal{O}_{ij, mn}({\bf \hat{k}}) \mathcal{O}_{jk, op}({\bf \hat{k}'}) \mathcal{O}_{ki, qr}({\bf \hat{k}}'')\delta({\bf k}+{\bf  k}'+{\bf k}'')\frac{(16\pi G)^{3}}{k k' k''} N\left(6\pi^2\frac{v^{2}}{N}\frac{\Gamma(3/2)\Gamma(7/2)}{\Gamma(1)}\right)^{3}I_{mnopqr}(k).
\end{eqnarray}
The integral is
\begin{eqnarray}\nonumber
I^{mnopqr}(k) & = & 
3! 2H_{0}^{3}\Omega_{\rm rad}^{3/2}\int_{\eta_{*}}^{1/k}d\tau_{1}\int_{\eta_{*}}^{\tau_{1}}d\tau_{2}\int_{\eta_{*}}^{\tau_{2}}d\tau_{3}\; k^{3}\tau_{1}^{5}\tau_{2}^{5}\tau_{3}^{5}\Big(\int_{0}^{1/\tau_{1}} dp\, I^{mnopqr}_{1}+\int_{1/\tau_{1}}^{1/\tau_{2}}dp\, I^{mnopqr}_{ 2}
\\ && \quad+\int_{1/\tau_{2}}^{1/\tau_{3}}dp\, I^{mnopqr}_{ 3}+\int_{1/\tau_{3}}^{1/k}dp\,I^{mnopqr}_{4} \Big),
\end{eqnarray}
where the factor of $3!$ accounts for permutations of the integration variables, $\{\tau_{1}, \tau_{2}, \tau_{3}\}$, and we have used the scale factor during radiation domination $a \approx H_{0}\sqrt{\Omega_{\rm rad}}\tau$, consistent with $a_{0} = 1$ today. The integrals are
\begin{eqnarray}
I^{mnopqr}_{ 1} & = & \frac{1}{(2\pi)^{3}2^{18}} \int d\Omega\; p^mp^np^op^p(p-k)^q(p-k)^r\\\nonumber
I^{mnopqr}_{2} & = & \frac{1}{(2\pi)^{3}2^{12}}\frac{2}{\pi}\frac{\cos^{2}\left(p\tau_{1}+\frac{5\pi}{4}\right)^{2}}{(p\tau_{1})^{5}} \int d\Omega\; p^mp^np^op^p(p-k)^q(p-k)^r\\\nonumber
I^{mnopqr}_{3} & = & \frac{1}{(2\pi)^{3}2^{6}}\frac{2}{\pi}\frac{\cos^{2}\left(p\tau_{1}+\frac{5\pi}{4}\right)^{2}}{(p\tau_{1})^{5}}\frac{2}{\pi}\frac{\cos^{2}\left(p\tau_{2}+\frac{5\pi}{4}\right)^{2}}{(p\tau_{2})^{5}} \int d\Omega\; p^mp^np^op^p(p-k)^q(p-k)^r\\\nonumber
I^{mnopqr}_{4} & = & \frac{1}{(2\pi)^{3}}\frac{2}{\pi}\frac{\cos^{2}\left(p\tau_{1}+\frac{5\pi}{4}\right)^{2}}{(p\tau_{1})^{5}}\frac{2}{\pi}\frac{\cos^{2}\left(p\tau_{2}+\frac{5\pi}{4}\right)^{2}}{(p\tau_{2})^{5}}\frac{2}{\pi}\frac{\cos^{2}\left(p\tau_{3}+\frac{5\pi}{4}\right)^{2}}{(p\tau_{3})^{5}} \int d\Omega\; p^mp^np^op^p(p-k)^q(p-k)^r
\end{eqnarray}
To perform these integrals, we replace the $\cos^{2}(p\tau)$ terms by their value averaged over a few cycles, $1/2$.  The angular integrals can be done in the usual way. In the equilateral limit\begin{eqnarray}\nonumber\label{eqn:psix}
\mathcal{O}_{ij, mn}({\bf \hat{k}}) \mathcal{O}_{jk, op}({\bf \hat{k}'}) \mathcal{O}_{ki, qr}({\bf \hat{k}}'')\int d\Omega\; \hat{p}^m\hat{p}^n\hat{p}^o\hat{p}^p\hat{p}^q\hat{p}^r  & = &  \frac{9\pi}{80},\\
k^{o}k^{p}\mathcal{O}_{ij, mn}({\bf \hat{k}}) \mathcal{O}_{jk, op}({\bf \hat{k}'}) \mathcal{O}_{ki, qr}({\bf \hat{k}}'')\int d\Omega\; \hat{p}^m\hat{p}^n\hat{p}^q\hat{p}^r & = &  -\frac{69\pi}{640}k^{2}.
\end{eqnarray}
Finally on subhorizon scales we obtain 
\begin{eqnarray} \label{eqn:GPTanswer}
\langle hhh\rangle_{\rm \Delta}\delta({\bf k}+{\bf k}'+{\bf k}'') & \sim & -\frac{20}{N^2}\frac{1}{k^6} \left(\frac{v}{M_{\rm pl}}\right)^{6}\left(\frac{H_{0}\sqrt{\Omega_{\rm rad}}}{k}\right)^{3}\delta({\bf k}+{\bf k}'+{\bf k}'').
\end{eqnarray}
The subscript $\Delta$ denotes the equilateral limit, $\Omega_{\mathrm{rad}}h^{2} = 4.15 \times 10^{-5}$ is the radiation density today and $N$ is the number of components in the scalar field, which is taken to be $\sim 4$.  

We can calculate the 2-pt function in the analogous way, using the propagator above in Eqn. (\ref{eqn:2ptlinear}) the integral is
\begin{eqnarray}\nonumber
\langle h_{ij}({\bf k }, \tau)h_{ij}({\bf k}', \tau)\rangle & = & N\left(6\pi^{2}\frac{v^{2}}{N}\frac{\Gamma(3/2)\Gamma(7/2)}{\Gamma(1)}\right)^{2}\mathcal{O}_{kl,mn}({\bf k})\frac{\left(16 \pi G\right)^{2}}{kk'} \int_{\tau_{i}}^{\tau_{f}}d\tau_{1}\int_{\tau_{i}}^{\tau_{f}}d\tau_{2}a(\tau_{1})\sin(k\tau_{1})a(\tau_{2})\sin(k\tau_{2})\\&&
\times\tau_{1}^{3}\tau_{2}^{3}\int\frac{d^{3}p}{(2\pi)^{3}}p_{k}p_{l}p_{m}p_{n}\frac{J_{2}(p\tau_{1})}{(p\tau_{1})^{2}}\frac{J_{2}(p\tau_{2})}{(p\tau_{2})^{2}}\frac{J_{2}(|{\bf p}-{\bf k}|\tau_{1})}{(|{\bf p}-{\bf k}|\tau_{1})^{2}}\frac{J_{2}(|{\bf p}-{\bf k}|\tau_{2})}{(|{\bf p}-{\bf k}|\tau_{2})^{2}}\delta({\bf k}+{\bf k}').
\end{eqnarray}
In the same approximation as above, we find for modes inside the horizon
\begin{eqnarray}\label{eqn:2ptfinal}
\langle h_{ij}({\bf k }, \tau)h_{ij}({\bf k}', \tau)\rangle \sim \frac{\pi}{N}\frac{1}{k^{3}}\left(\frac{v}{M_{\rm pl}}\right)^{4}\left(\frac{H_{0}\sqrt{\Omega_{\rm rad}}}{k}\right)^{2}\delta({\bf k}+{\bf k}').
\end{eqnarray}
Then, writing 
\begin{eqnarray} \label{eqn:niceeqn}
k^{6}\langle  hhh \rangle_{\Delta}  = \mathcal{C}_{NL}(k^{3}\langle hh \rangle)^{3/2},
\end{eqnarray}
where $\mathcal{C}_{NL}$ is a dimensionless constant, for this theory we have 
\begin{equation}
\mathcal{C}_{NL} =\frac{20}{\pi^{3/2}\sqrt{N}}\sim \mathcal{O}(1).
\end{equation}
for $N=4$. For inflation $\mathcal{C}_{NL}\sim \mathcal{O}(H/M_{\rm pl})$, while for this model the 3-pt function is as important as the 2-pt function. Note that $\mathcal{C}_{NL}$ is \emph{independent} of the scale at which the process is occurring. This means that, even if a global phase transition can completely mimic inflation at the level of the power spectrum, it produces a 3-pt which is distinct from and much larger than that of inflation. 
The existence of such a large 3-pt correlation function relative to the 2-pt correlation function suggests that if a B-mode signal consistent with $r\sim 0.01$ was observed, by constructing the 3-pt estimator we might determine whether its origins were consistent with inflation. 

We also point out that this 3-pt would vanish in direct detection experiments for precisely the same reasons as in the preheating case. At first it may seem that since the gravitational waves are continuously sourced that we might be able to evade the simple argument above based on the central limit theorem. However, the power at each scale, $k$, is sourced as that particular mode enters the horizon. Since direct detection experiments are sensitive to scales on the order of the size of the solar system, the gravitational waves detected in these experiments will be primarily composed of radiation which was emitted when the horizon scale was on the order of the size of our solar system. This means that these gravitational waves will again look almost completely Gaussian by the central limit theorem.

\section{Conclusions and Future Outlook}\label{sect:conclusions}

In this paper we considered the properties of the 3-point statistics of cosmological gravitational waves from both inflationary and non-inflationary ``active'' scalar sources. For the latter, we consider gravitational waves  from preheating driven turbulence at the end of inflation and from self-ordering scalar fields following a global phase transition in the early universe.

Introducing a ``3-pt correlation parameter'' ${\cal C}_{NL}$, we write
\begin{equation}
k^{6}\langle  hhh \rangle_{\Delta}  = \mathcal{C}_{NL}(k^{3}\langle hh \rangle)^{3/2},
\end{equation}
where the $_\Delta$ here denotes the equilateral limit of the bispectrum. During inflation, metric fluctuations (gravitons) start in a purely Gaussian state and interactions with other fluctuations are  highly suppressed by the amplitude of the observed scalar spectrum. This means that the leading order effective three graviton interaction is the tree level interaction, and any departure from Gaussianity is highly suppressed. For GUT scale inflation, ${\cal C}_{NL} \sim H_{\rm inf}/M_{\rm Pl}\sim 10^{-6}$, and ${\cal C}_{NL}$ measures the the scale of inflation directly, which provides a consistency check when compared to the SGW power spectrum.  Unfortunately, and unsurprisingly, we find that even a BBO/DECIGO class detector cannot detect an inflationary 3-pt signal.

On the other hand, for SGW from scalar sources, ${\cal C}_{NL}$ measures the strength of the \emph{effective} gravitational wave interaction, i.e. it is mediated by scalars. For these sources there is no reason to expect the higher point functions to be suppressed relative to the 2-pt function. In highly inhomogenous phenomenon that follows any period of cosmological turbulence, we estimate that ${\cal C}_{NL} \sim {\cal O}(1) \gg  H_{inf}/M_P$. 

We investigated two different scalar sources of SGW. The first model is that of gravitational waves produced in the era of preheating after inflation. During preheating, gravitational waves are produced with a characteristic wavelength $k_p \sim H_{e}$ where $H_{e}$ is the Hubble scale at the end of inflation. While the gravitational waves are highly correlated at this scale, since $H_{e} \gg H_{today}$, such sources appear as \emph{uncorrelated} patches in the sky. That is, these patches appear to our detectors as uncorrelated point sources. The gravitational radiation we observe in our detectors is the sum of gravitation waves coming from all directions, and is thus composed of gravitational waves from many patches. By constructing the correlator of three time streams, we are integrating the estimator $\langle h({\bf n}, f)h({\bf n}', f')h({\bf n}'', f'') \rangle$ over the entire sky. Clearly the 3-pt signal from such a source will be highly Gaussian via the central limit theorem -- ironically even more so than the inflationary signal. Nevertheless, the gravitational radiation from processes such as TeV scale preheating is expected to be easily seen by future direct detection experiments. The high level of Gaussianity of these stochastic backgrounds then presents the intriguing possibility of using them as a probe of the intervening matter distribution.

The second model we consider is gravitational raditation from self ordering scalar fields following a global phase transition proposed in this context by Jones-Smith, Krauss and Mathur \cite{JonesSmith:2007ne}. In this model, a scale invariant spectrum of SGW is generated  by self ordering scalar fields at the horizon. We show that the correlation parameter ${\cal C}_{NL} \sim {\cal O}(1)$ in this model and a large 3-pt correlation is expected.  Nevertheless, since these gravitational waves are being continuously sourced at the horizon, the central limit argument above still applies and the 3-pt signal today will be small to vanishing.

One potential place where a non-Gaussian gravitational wave signal may still be detected is in the polarization measurements of the CMB. The gravitational waves sourced by the global phase transition mechanism are completely correlated, which means that we expect that as well as the usual B-mode spectrum, $\langle BB \rangle$, they will also give rise to non-trivial $\langle BBB\rangle$ and  $\langle EBB\rangle$ correlations, and could provide an independent discriminant of non-inflationary sources of $B$ mode polarizations.

\acknowledgements

We thank Mustafa Amin, Niayesh Ashfordi,   Latham Boyle, Alessandra Buonanno, Richard Easther, Raphael Flauger, Lawrence Krauss, Ue Li Pen and Harsh Mathur and  for many helpful discussions. We would especially like to thank Lam Hui and Daniel Baumann for many insightful comments, and for pointing out some crucial errors in our original draft. EAL would like to thank the Chinese Academy of Sciences and the KITP(China) for hospitality where some of this work was done. PA is supported in part by the United States Department of Energy, grant DE-FG02-92ER-40704 and by NSF grant PHY-0747868.  EAL is supported in part by the DOE  (DE-FG02-92-ER40699) and in part by the Project of Knowledge Innovation Program (PKIP) of Chinese Academy of Sciences, Grant No. KJCX2.YW.W10

\appendix

\section{The 3-pt Estimator for Direct Detection Experiments }  \label{sect:3pt}
In this Appendix, we calculate the correlation of three streams of data from a direct detection experiment such as LISA. A detector output is a scalar stream of data $S(t)$ as a function of time, where the data measure the direct strain of the detector. The data stream consists of a signal $s(t)$ and an inherent noise $n(t)$
\begin{equation}
S(t) = s(t) + n(t).
\end{equation}
In general we are deep in the noise dominated regime, $|n| \gg |s|$, $\langle s\rangle = \langle n \rangle = 0$, and for two detectors in different locations, the noise is assumed to be localized, $\langle n_{1}n_{2}\rangle \ll \langle n_{1}n_{1}\rangle$, $\langle n_{2}n_{2}\rangle$. In the measurement of a stochastic \emph{power} spectrum, we can correlate two detectors to extract the signal out of the noise. In this case, the signal to noise ratio for the stochastic power spectrum is
\begin{equation}\label{eqn:2ptSNR}
\mathrm{SNR} \sim \frac{\langle S_{12}\rangle}{\sqrt{\langle N_{12}^{2} \rangle}},
\end{equation}
where $\langle S_{12}\rangle$ is the expectation value of the signal
\begin{eqnarray}
S_{12} = \int_{-T/2}^{T/2}dt_{1}\int_{-T/2}^{T/2}dt_{2}S_{1}(t_{1})S_{2}(t_{1})Q(t_{1}, t_{2}),
\end{eqnarray}
and $Q(t_{1}, t_{2})$ is a 2-pt filter function. The denominator of Eqn. (\ref{eqn:2ptSNR}) is the root mean square noise $\sqrt{\langle N_{12}^{2} \rangle}$, where $N_{12} = S_{12}-\langle S_{12}\rangle$. Since the noise is localized, $\langle S_{12}\rangle = \int dt_{1}\int dt_{2}\langle s_{1}(t_{1})s_{2}(t_{2}) \rangle W(t_{1}, t_{2}) $. 

By choosing a filter with a moving window, $Q(t_{1},t_{2}) = Q(t_{1}-t_{2})$, we see that the signal to noise ratio in Eqn. (\ref{eqn:2ptSNR}) scales like $\sqrt{T}$, where $T$ is the total integration time. The signal in the numerator is correlated and hence scales like $T$ while the denominator is uncorrelated, and so increases like a one dimensional random walk, $\sqrt{T}$.

With three detectors, we can construct the analogous estimator for the 3-pt using a filter $W(t_1,t_2,t_3)$,
\begin{eqnarray}\label{eqn:3ptestimator}
S_{123} & = & \int_{-T/2}^{T/2}dt_{1}\int_{-T/2}^{T/2}dt_{2}\int_{-T/2}^{T/2}dt_{3}S_{1}(t_{1})S_{2}(t_{2})S_{3}(t_{3})W(t_{1}, t_{2},t_{3}).
\end{eqnarray}
Localization of the detector noise means that 
\begin{eqnarray}\label{eqn:3ptsigexpect}
\langle S_{123} \rangle = \int_{-T/2}^{T/2}dt_{1}\int_{-T/2}^{T/2}dt_{2}\int_{-T/2}^{T/2}dt_{3}\langle s_{1}(t_{1})s_{2}(t_{2})s_{3}(t_{3})\rangle W(t_{1}, t_{2},t_{3}).
\end{eqnarray}
The noise here is $N_{123} = S_{123} - \langle S_{123}\rangle$, and the signal to noise ratio is
$
\mathrm{SNR}\sim \langle S_{123}\rangle/\sqrt{\langle N_{123}^{2}\rangle }.
$

To observe a signal at a frequency $f_{*}$ we must make a measurement that is at least $\Delta T \sim 1/f_{*}$ in length. Then, instead of making one long measurement over the total time $T$, we chop our signal into ``chunks'' of length $\Delta T \sim 1/f_*$, where $f_*$ corresponds to the characteristic minimal noise frequency of a given detector, i.e. $N_a(f_*) = \mathrm{minimal}$, where the subscript $a$ here labels the detector. We then have $T/\Delta T$ chunks of identical measurement of the signal $\langle s_{123}\rangle_M$, where $M$ labels the $M$-th chunk. On the other hand, the noise is uncorrelated across chunks and, as noted above, increases like a one-dimensional random walk.  The signal to noise per chunk is then 
$
\mathrm{SNR}_{M} = \langle S_{123}\rangle_M/\sqrt{N_{123}^2}$, where $\langle S_{123}\rangle_M$ is the expectation of Eqn. (\ref{eqn:3ptestimator}) after an integration time $\Delta T \sim 1/f_* $.

The \emph{total} SNR is then given by
\begin{equation}\label{eqn:3ptSNR}
\mathrm{SNR} = \frac{\langle s_{123}\rangle_M}{\sqrt{N_{123}^2}}  \times \sqrt{M} \propto \sqrt{T}.
\end{equation}
That is, the signal to noise ratio of a 3-pt correlator also scales like $\sqrt{T}$. This means that a detection is inevitable as long as we integrate for long enough (and the 3-pt is actually non-zero).

We now need to relate the estimator in Eqn. (\ref{eqn:3ptestimator}) above to the predicted 3-pt signal, $\langle h_{ij}({\bf x}, t_{1})h_{jk}({\bf x}',t_{2})h_{ki}({\bf x}'', t_{3})\rangle$ where the Roman indices run over the 3 spatial dimensions. We can expand any massless,  transverse-traceless tensor mode, $h_{ij}({\bf{x}},t)$, as 
\begin{equation}
h_{ij}({\bf x },t) = \int d\dho \int f^{2}\,df \sum_{A} h(\dho,f)e^{-2\pi i f(t-\dho \cdot {\bf x})}e_{ij}^A(\dho),
\end{equation}
where ${\bf k} = 2\pi f \dho~,~ k = 2\pi f$ and the speed of light $c=1$.  Our convention differs from that of the gravitational wave community, who usually absorb the $f^{2}$ into the amplitude  (see for example \cite{Allen:1996vm, Maggiore:1999vm}), but conforms with standard analytical techniques used to compute higher order correlation functions.
The polarization tensors are normalized via
\begin{equation}
e_{ij}^A({\bf k})e_{ij}^{A'}({\bf k}) = 2\delta^{AA'},
\end{equation}
where polarization is indexed with $A$. The explicit form of the polarization tensors can be found in (for example) \cite{Maggiore:1999vm}.

A gravitational wave is a spin-2 tensor field propagating through space at the speed of light. To detect a gravitational wave, it has to interact with a detector such as an interferometer or resonant mass. Each detector, labeled $a$, has a pattern tensor $D_{a}^{ij}$, which measures its response to a passing GW as a function of its geometry. We define the detector pattern function as
\begin{equation}
D^{ij}_{a}e^A_{ij}(\dho) = F^A_a(\dho).
\end{equation}
A detector located at ${\bf x}_{a}$ then produces a  data stream  $S_a(t, {\bf x}_{a}) = s_{a}(t, {\bf x}_{a})+n_{a}(t, {\bf x}_{a})$ in response to a passing GW, where $s_{a}(t, {\bf x}_{a})$ is
\begin{equation}
s_a(t, {\bf x}_{a}) = \int f^{2}\,df d\dho \sum_A h^A(f,\dho)e^{-2\pi i f(t-\dho \cdot {\bf x}_{a})}F^A_a(\dho) \label{eqn:signalexpand}.
\end{equation}

With three data streams $S_1(t_{1})$, $S_2(t_{2})$ and $S_3(t_{3})$, and imposing the filter function $W(t,t',t'')$, the expectation of the estimator $\langle S_{123}\rangle$ in Eqn. (\ref{eqn:3ptsigexpect}) is related to the GW 3-pt function by
\begin{eqnarray}\nonumber
\langle S_{123} \rangle &=& \int^{T/2}_{-T/2} dt\int^{T/2}_{-T/2} dt'\int^{T/2}_{-T/2} dt'' d\dho d\dho' d\dho'' f^{2}\,df f'^{2}\, df' f''^{2}\,df'' \sum_{A,A',A''}  \langle h^A(f,\dho) h^{A'}(f',\dho')h^{A''}(f'',\dho')\rangle \\
&&\qquad\qquad \times F_1^A(\dho) F_2^{A'}(\dho')F_3^{A''}(\dho'') e^{-2\pi i[f(t - \dho \cdot {\bf x}_1)+f'(t'- \dho' \cdot {\bf x}_2)+f''(t'' - \dho'' \cdot {\bf x}_3)]}W(t,t',t'') \label{eqn:master}.
\end{eqnarray}
Eqn. (\ref{eqn:master}) is the master formula for the signal. Given a predicted 3-pt correlation function and detector setup we can calculate the expected signal to noise.

Taking the Fourier transform of the window function, $W(t,t', t'')$, and breaking up our total integral into chunks of length $\Delta T > 1/f_{*}$ we can perform the time integrals\footnote{The results of the time integrations are really truncated delta functions. We ignore this here for the purposes of estimation.}  to obtain the signal per chunk
\begin{eqnarray}\nonumber\label{eqn:signalm}
\langle S_{123} \rangle_M
 &=& \int d\dho d\dho' d\dho'' f^{2}df f'^{2} df' f''^{2}df'' \sum_{A,A',A''}  \langle h^A(f,\dho) h^{A'}(f',\dho')h^{A''}(f'',\dho')\rangle \\
&&\qquad\qquad \times F_1^A(\dho) F_2^{A'}(\dho')F_3^{A''}(\dho'') e^{2\pi i(f \dho \cdot {\bf x}_1+f' \dho' \cdot {\bf x}_2+f'' \dho'' \cdot {\bf x}_3)}\tilde{W}(f,f',f''). \label{eqn:signal123}
\end{eqnarray}

In general, the signal is reduced by the factors $F^{A}_a(\dho)$ and  $\exp[2\pi i(f \dho \cdot {\bf x}_1+f' \dho' \cdot {\bf x}_2+f'' \dho'' \cdot {\bf x}_3)]$.  $F_a^A(\dho)\leq  1$ because the detectors generally do not possess isotropic beams in general and detectors are often not optimally aligned. The relative location exponent, $\exp[2\pi i(f \dho \cdot {\bf x}_1+f' \dho' \cdot {\bf x}_2+f'' \dho'' \cdot {\bf x}_3)]\leq1$, as detectors that are not co-located can destructively interfere. This is because detectors in different locations are measuring different parts of the wave. In the calculation of the 2-pt correlator, the angular integral over these factors is collectively called the \emph{overlap reduction function} \cite{Flanagan:1993ix}, $\Gamma(f)$, and encodes the effect of the detectors on the correlation. In the case of the 3-pt there is no easy distillation of this information into a simple factor (like $\Gamma(f)$) because in Eqn. (\ref{eqn:master}), the correlation function itself involves a triangle constraint in momentum space $\delta(f\dho+f'\dho'+f''\dho'')$ which makes separation of the angular information difficult.

As noted above, localization means that the noise component of the signal above is zero. The noise is then the variance of the above estimator $N_{123} = S_{123} - \langle S_{123} \rangle$. By definition $\langle N \rangle = 0$, and we take $N  = \sqrt{\langle N_{123}^{2}\rangle} $. Defining as usual (\cite{Allen:1996vm, Maggiore:1999vm})
\begin{eqnarray}
n_{a}(t) & = & \int df n_{a}(f)e^{-2\pi i f},\quad\textrm{with}\quad \langle n_{a}(f)n_{b}(f')\rangle = \frac{1}{2}\delta_{ab}\delta(f-f')N_{a}^{2}(f),
\end{eqnarray}
where the indices $a,b$ label the detector then the noise is
\begin{equation}
\langle N_{123}^2 \rangle = \int df df' df''  \frac{1}{8}N_1(f)^2N_2(f')^2N_3(f'')^2 |\tilde{W}(f,f',f'')|^2 .\label{eqn:noisegeneral}
\end{equation}
Given a 3-pt correlator, Eqns (\ref{eqn:signalm}) and (\ref{eqn:noisegeneral}) are the ingredients we need to estimate detection sensitivity.

Evaluating the expression for the signal in Eqn. (\ref{eqn:signalm}) involves messy calculations and requires knowledge of the detector set-up, locations and orientations. Given a signal and detector configuration, an \emph{optimal} filter can be designed. In principle, each different 3-pt correlation function will yield a different optimal filter. For our purposes, this is an unnecessary level of detail. Instead,  we will make several simplifying assumptions in order to  estimate the signal to noise ratio:

\begin{itemize}

\item{We assume isotropic detectors, so $F^{A}(\dho) =1~ \forall~ (\dho,A) $. In general, these factors can be small to vanishing if our detectors are misaligned.}

\item{We assume co-located detectors, so $\exp(f \dho \cdot {\bf x}_1+f'\dho'\cdot {\bf x}_2+ f''\dho'' \cdot {\bf x}_3) =1 $ via  momentum conservation.}

\item{We use an idealized noise spectrum, where the noise $N(f)$ is an inverse tophat with minimum $N(f_*)$ in the domain $(f_* -\Delta f,f_*+ \Delta f)$.  We then choose simple tophat filters $\tilde{W}=\tilde{W}_1(f)\tilde{W}_2(f)\tilde{W}_3(f)$, such that $\tilde{W}_1(f) = \tilde{W}_2(f) = \tilde{W}_3(f) = \theta(f_*+\Delta f - f)\theta(f-f_*+\Delta f)$, where $f_*$ is the optimal noise frequency, i.e. the filter has support for the frequency range of $2\Delta f$ around this optimal frequency. Clearly this filter is suboptimal -- the optimal filter depends on both the exact noise spectra, the configuration of the detectors and the actual signal itself.}

\end{itemize}

\begin{figure}
\centering
\includegraphics[scale = 0.6]{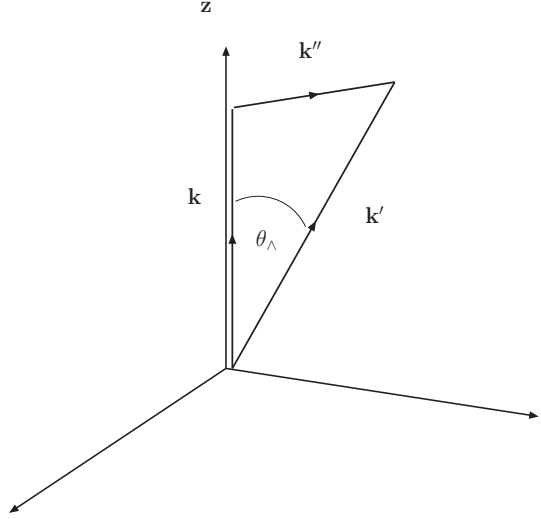}\caption{The 3 momenta must form a triangle via conservation. By assuming isotropy of both signal and detector configuration (as we have done here), the integral Eqn. (\ref{eqn:signal123}) can be reduced to an integral over all possible shapes of a triangle as shown in this figure. The measure is then completed by rotating around the $z$ axis, and then around all possible directions of its axis, for a total of $8\pi^2$.}\label{fig:triangles}
\end{figure}

Using these assumptions, we can simplify the integration of Eqn. (\ref{eqn:signalm}). The delta function imposes a triangle condition on the momenta, defining a plane in the 3 dimensional momentum space. The signal is assumed to be isotropic, and so without any loss of generality, we can pick $\dho$ to point in the $z$ direction. Focusing on the triangles defined on the planes orthogonal to this direction (see fig. \ref{fig:triangles}),
%
%
the integral sums over all the possible shapes of a single triangle.  We write
\begin{equation}
d\dho d\dho' d\dho'' {\cal F}(f,f,f'') \delta (f\dho + f'\dho'+f''\dho'') \rightarrow  d\cos \theta_{\wedge} 8\pi^2 {\cal F}(f,f,f'')\frac{1}{f^2}\delta(f-\sqrt{f'{}^2+f''{}^2+2f'f''\cos \theta_{\wedge}}),
\end{equation}
where $\theta_{\wedge}$ is the angle between $\dho'$ and $\dho''$ and ${\cal F}(f, f', f'')$ is defined by 
\begin{eqnarray}
\langle h(f, \dho)h(f', \dho')h(f'', \dho'')\rangle = \mathcal{F}(f,f',f'')\delta(f\dho+f'\dho'+f''\dho'').
\end{eqnarray}
The factor of $8\pi^2$ is the measure obtained by revolving the triangle around the $z$ axis ($2\pi$), and then over the solid angle spanned by $\dho$ ($4\pi$), as allowed by isotropy. The tophat filters then constrain the shapes of triangles which contribute to the signal, hence the range of $\theta_{\wedge}$. It is easy to see that this implies 
\begin{equation}
\cos \theta_{\wedge}^{max} = \frac{(f_*+\Delta f)^2 -2(f_*- \Delta f)^2}{2(f_* - \Delta f)^2},\quad \textrm{and}\quad
\cos \theta_{\wedge}^{min} = \frac{(f_*-\Delta f)^2 -2(f_*+ \Delta f)^2}{2(f_* + \Delta f)^2}.
\end{equation}
Eqn. (\ref{eqn:signalm}) becomes
\begin{equation}
\langle S_{123}\rangle_M= \int^{\theta_{\wedge}^{max}}_{\theta_{\wedge}^{min}} d\cos \theta_{\wedge} \int_{f_*-\Delta f}^{f_*+\Delta f} f^{2}df f'^{2}df' f''^{2} df''  {\cal F}(f,f,f'')  \frac{8\pi^2}{f^2}\delta(f-\sqrt{f'{}^2+f''{}^2+2f'f''\cos \theta_{\wedge}}).
\label{eqn:reducedmaster}
\end{equation}
The simple noise model means the 3-pt noise is given by
\begin{equation}
\langle N_{123}^2\rangle = N^{2}_1(f_*)N^{2}_2(f_*)N^{2}_3(f_*) (\Delta f)^3.
\end{equation}

Before we press on to calculate the prospects for detection, compare the estimator Eqn. (\ref{eqn:reducedmaster}) to the more familiar 2-pt SNR. The signal to noise for the 3-pt scales as $(\Delta f)^{3/2}$ while for the 2-pt at first approximation do not scale with $(\Delta f)$ -- hence the 3-pt SNR is more susceptible to sampling rate effects. As long as we keep the sampling rate $X\gg f_*$, this effect is negligible.

\section{SGW from scalar sources} \label{app:scalarsources}

In this work, we follow \cite{Dufaux:2007pt, Price:2008hq}, and present some basic results regarding the generation of a 3-pt correlation function of stochastic gravitational waves by cosmological scalar fields. The key results are Eqns (\ref{eqn:full3pt}) and its linear field limit, Eqn. (\ref{eqn:gaussianpreheat}).

The results we present here are completely general, and apply to any process in which a scalar field evades the no-go theorem of Dufaux et. al. \cite{Dufaux:2007pt}. Examples of such processes are those considered above, preheating and global phase transitions. Since preheating is a highly non-linear, non-perturbative phase, numerical simulations and evaluation of Eqn. (\ref{eqn:full3pt}) are required for accurate results. This is outside the scope of this paper. During the early part of reheating and during global phase transitions, however, the fields remain linear and thus it is possible to calculate analytically. In this limit we obtain Eqn. (\ref{eqn:gaussianpreheat}). We evaluate this expression approximately for the case of the global phase transition of \cite{JonesSmith:2007ne, Fenu:2009qf}. A similar result can be obtained for preheating using the technology of \cite{Dufaux:2007pt, Price:2008hq}, however, one must numerically simulate occupation numbers of the preheating fields, and we leave it for future work.

As noted above, gravitational wave production by scalar fields in a non-inflationary regime is substantially different from gravitational wave production during inflation. Quantum effects are negligible and purely classical effects, e.g. the relaxation of a disordered field or the turbulent motion of large masses, lead to the emission of gravitational waves. In this case, gravitational waves are sourced by the transverse traceless part of the anisotropic stress $\Pi_{ij}$
\begin{eqnarray}
\Pi_{ij} & = & T_{ij}- \langle p\rangle g_{ij}.
\end{eqnarray}
where the energy momentum tensor for the inhomogeneous scalar fields is the usual
\begin{eqnarray}
T_{\mu\nu} & = & \partial_{\mu}\phi_{a}\partial_{\nu}\phi_{a}-g_{\mu\nu}\left(\frac{1}{2}g^{\alpha\beta}\partial_{\alpha}\phi_{a}\partial_{\beta}\phi_{a}+V\right).
\end{eqnarray}
The equation of motion for the gravitational waves in an expanding background with the above source is
\begin{eqnarray}
\bar{h}_{ij}''({\bf k},\tau)+\left(k^{2}-\frac{a''}{a}\right)\bar{h}_{ij}({\bf k},\tau) & = & 16 \pi G a^{3} \Pi^{\rm TT}_{ij}({\bf k})
\end{eqnarray}
where a prime, $'$, denotes a derivative with respect to conformal time and $\bar{h}_{ij} = a h_{ij}$ is the comoving metric amplitude. The traceless, transverse part of the anisotropic stress is found using the transverse traceless projector
\begin{eqnarray}\label{eqn:transversetraceless}
\Pi^{\rm TT}_{ij}({\bf k}) & = & \mathcal{O}_{ij,lm}({\bf \hat{k}})\Pi_{lm}({\bf k}) = \left[P_{il}({\bf \hat{k}})P_{jm}({\bf \hat{k}}) - \frac{1}{2}P_{ij}({\bf \hat{k}})P_{lm}({\bf \hat{k}})\right]\Pi_{lm}({\bf k})
\end{eqnarray}
where the projector is $P_{ij}({\bf \hat{k}}) =  \delta_{ij} - \hat{k}_{i}\hat{k}_{j}$. Now, as pointed out by \cite{Dufaux:2007pt} the only relevant part of the energy momentum tensor is the product of the spatial derivatives, so
\begin{eqnarray}\nonumber
a^{2}\Pi_{ij}^{\rm TT}({\bf k}) & = & \left[P_{il}({\bf \hat{k}})P_{jm}({\bf \hat{k}}) - \frac{1}{2}P_{ij}({\bf \hat{k}})P_{lm}({\bf \hat{k}})\right] \int \frac{d^{3}p}{(2\pi)^{3}}\int \frac{d^{3}p'}{(2\pi)^{3}}p^{l}p'^{m}\phi_{a}({\bf p})\phi_{a}({\bf p'})\delta({\bf k}-{\bf p} -{\bf p'})\\
& \equiv & T_{ij}^{\rm TT}({\bf k})
\end{eqnarray}
Now, since the gravitational waves originate from relatively short intervals of time, in this work we neglect the expansion of space, dropping the $a''/a$ factor
\begin{eqnarray}
v''_{k}\epsilon_{ij}+k^{2}v_{k}\epsilon_{ij} & = & 16 \pi G\; a(\tau) T_{ij}^{\rm TT}({\bf k})
\end{eqnarray}
where $\bar{h}_{ij}({\bf k}, \tau) = v_{k}(\tau)\epsilon_{ij}({\bf k})$. We can construct the Green's functions for this Eqn. \cite{Dufaux:2007pt, Baumann:2007zm}
\begin{eqnarray}\label{eqn:greenfn}
G(\tau; \tau') 
 & = & \frac{1}{k}\sin(k(\tau-\tau'))
\end{eqnarray}
then,
we have
\begin{eqnarray}
 \bar{h}_{ij}(\tau, {\bf k})  & = &  \frac{16 \pi G}{k} \int_{\tau_{i}}^{\tau} d\tau'\sin(k(\tau-\tau'))a(\tau')T_{ij}^{\rm TT}({\bf k}, \tau'),\\
\bar{h}'_{ij}(\tau, {\bf k}) & = & 16 \pi G\int_{\tau_{i}}^{\tau} d\tau'\cos(k(\tau-\tau'))a(\tau')T_{ij}^{\rm TT}({\bf k}, \tau').
\end{eqnarray}
Now, supposing that the source is turned off at some time $\tau_{f}$, then the gravitational waves becoming freely propagating, hence the solution becomes
\begin{eqnarray}
h_{ij}({\bf k}, \tau) & = & \frac{v_{k}(\tau)}{a(\tau)}\epsilon_{ij}({\bf k})   =  A_{ij}({\bf k})\frac{\sin[k(\tau - \tau_{f})]}{a(\tau)}+B_{ij}({\bf k})\frac{\cos[k(\tau - \tau_{f})]}{a(\tau)},
\end{eqnarray}
where $A_{ij}({\bf k})$ and $B_{ij}({\bf k})$ are obtained from matching the solution at $\tau = \tau_{f}$;
\begin{eqnarray}
A_{ij}({\bf k}) & = & \frac{16\pi G}{k}\int_{\tau_{i}}^{\tau_{f}}d\tau' \cos[k(\tau_{f}-\tau')]a(\tau')T_{ij}^{\rm TT}({\bf k}, \tau'),\\
B_{ij}({\bf k}) & = & \frac{16\pi G}{k}\int_{\tau_{i}}^{\tau_{f}}d\tau' \sin[k(\tau_{f}-\tau')]a(\tau')T_{ij}^{\rm TT}({\bf k}, \tau').
\end{eqnarray}
Now, today for frequencies $f \gg H_{0}$ we can Fourier transform, taking $a_{0} = 1$ we find, in real space 
\begin{eqnarray}
h_{ij}({\bf x}, t) & = & \int_{-\infty}^{\infty}f^{2} df \int d\Omega e^{-2\pi i f t}e^{ 2\pi  i k \tau_{f}}\left(A_{ij}({\bf{k}})\sin({\bf k}\cdot{\bf x})+B_{ij}({\bf{k}})\cos({\bf k}\cdot{\bf x})\right).
\end{eqnarray}
Working in the co-located approximation, we can choose ${\bf x} = 0$. We can now calculate the 3-pt function
\begin{eqnarray}\label{eqn:preheat3pt}
&&\langle h_{ij}({\bf k}, f)h_{jk}({\bf k'},f')h_{ki}({\bf k''},f'')\rangle \\\nonumber& = &   \frac{(16\pi G)^{3}}{k k' k''}\int_{\tau_{i}}^{\tau_{f}}d\tau d\tau'd\tau'' a(\tau)\sin[k\tau]a(\tau')\sin[k'\tau'] a(\tau'')\sin[k''\tau'']\left\langle T_{ij}^{\rm TT}({\bf k}, \tau)T_{jk}^{\rm TT}({\bf k}', \tau')T_{ki}^{\rm TT}({\bf k}'', \tau'')\right\rangle,
\end{eqnarray}
which leads us to consider the unequal time correlation function of three copies of the transverse traceless energy momentum tensor;
\begin{eqnarray}\nonumber\label{eqn:full3pt}
\langle T^{\rm TT}_{ij}({\bf k}, \tau)T^{\rm TT}_{jk}({\bf k}', \tau')T^{\rm TT}_{kl}({\bf k}'', \tau'')\rangle & = & \mathcal{O}_{ij, mn}({\bf \hat{k}}) \mathcal{O}_{jk, op}({\bf \hat{k}'}) \mathcal{O}_{ki, qr}({\bf \hat{k}}'')\int\frac{d^{3}p}{(2\pi)^{3}}\int \frac{d^{3}q}{(2\pi)^{3}}\int \frac{d^{3}s}{(2\pi)^{3}}p_{m}p_{n}q_{o}q_{p}s_{q}s_{r}\\
&&\Big\langle \phi_{a}({\bf p})\phi_{a}({\bf k}-{\bf  p})\phi_{b}({\bf q})\phi_{b}({\bf k}'-{\bf  q})\phi_{c}({\bf s})\phi_{c}({\bf k}''-{\bf  s})\Big\rangle,
\end{eqnarray}
where we have dropped terms which vanish (by transversality of $h_{ij}({\bf k})$).

\subsection{Linear Field Limit}

\begin{figure}
\centering
\includegraphics[scale = 0.6]{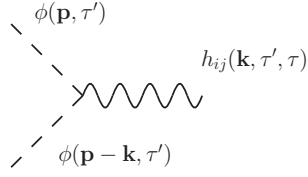}\caption{Vertex for generation of gravitational waves from linear scalars. Wiggly lines denote gravitational wave Green's functions while dashed lines denote scalar propagators (stochastic averages). \label{fig:vertex}}
\end{figure}
\begin{figure}
\centering
\includegraphics[scale = 0.6]{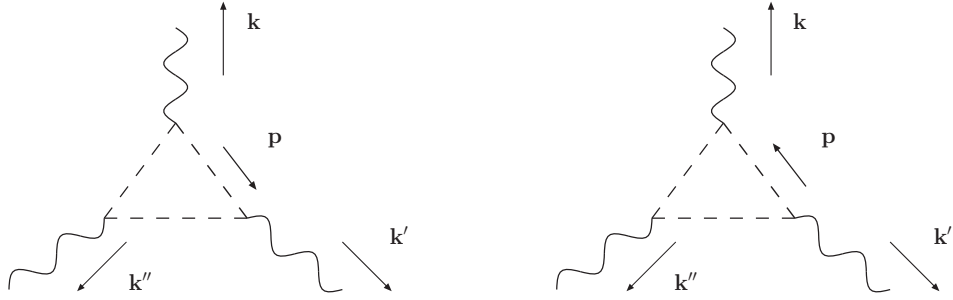}
\caption{Unlike inflation, SGW sourced by scalars (or any active sources) are generated at 1-loop order. Each vertex is of order $(h \partial_i \phi\partial_j \phi)^{TT}$ where TT indicates the traceless transverse part of the spatial partial derivatives on $\phi$, i.e. it is simply gravitational bremsstrahlung. The power spectrum generated by such a process will be a 1-loop diagram with two external GW legs. \label{fig:loopy}}
\end{figure}

In the limit that the scalar fields are well described by Gaussian statistics, we can then evaluate the 6-pt function of the scalar fields using Wick's theorem and the scalar field propagator \cite{Dufaux:2007pt}
\begin{eqnarray}
\langle \phi_{a}({\bf k}, \tau)\phi_{b}({\bf k'}, \tau')\rangle & = & F_{ab}(k, \tau, \tau')\delta({\bf k}+{\bf k'}).
\end{eqnarray} 
In this limit, one can understand the process in terms of the graph in fig \ref{fig:loopy}. Using the rules, 
\begin{enumerate}
\item Draw all diagrams, and label each vertex with a time.

\item A vertex as in fig. \ref{fig:vertex} gets a factor of $-16\pi G \mathcal{O}_{ij,mn}({\bf k})p_{m}p_{n}$, where $\mathcal{O}({\bf k})$ is the transverse traceless projector defined in Eqn. (\ref{eqn:transversetraceless})

\item An external graviton (gravitational wave) line gets a factor of the Green's function solution $G_{k}(\tau, \tau')$ to Eqn. (\ref{eqn:greenfn}), where $\tau$ is the time at which the diagram is being evaluated and $\tau'$ is the time associated with the vertex.

\item An internal scalar line gets a ``propagator,'' which is technically a stochastic average:
$\langle \phi_{a}({\bf p}, \tau)\phi_{b}({\bf p}, \tau') = F_{ab}(p, \tau, \tau')$

\item Conserve external momenta with an overall delta, $\delta(\sum_{i}{\bf k}_{i})$

\item Integrate over internal (loop) momenta

\item Integrate over times associated with each vertex from the initial time (when the interaction begins) to the final time when the source is turned off.

\end{enumerate}

We can evaluate the diagrams. The three point function is, after fourier transforming into frequency domain and using the co-located detector approximation
\begin{eqnarray}\label{eqn:gaussianpreheat}
&&\langle h_{ij}( f, \hat{\Omega})h_{jk}(f',\hat{\Omega}')h_{ki}(f'',\hat{\Omega}'')\rangle \\\nonumber& = &   -4\mathcal{O}_{ij, mn}({\bf \hat{k}}) \mathcal{O}_{jk, op}({\bf \hat{k}'}) \mathcal{O}_{ki, qr}({\bf \hat{k}}'')\delta({\bf k}+{\bf  k}'+{\bf k}'')\frac{(16\pi G)^{3}}{k k' k''}\int_{\tau_{i}}^{\tau_{f}}d\tau_{1} d\tau_{2}d\tau_{3} a(\tau_{1})\sin[k\tau_{1}]a(\tau_{2})\sin[k'\tau_{2}] a(\tau_{3})\sin[k''\tau_{3}]\\\nonumber
&&\qquad\qquad\times\int\frac{d^{3}p}{(2\pi)^{3}}p^{m}p^{n}\Big(p^{o}p^{p}(p-k)^{q}(p-k)^{r}F_{cb}(|{\bf k}' + {\bf p}|, \tau_{3},\tau_{2} ) F_{ab}({p}, \tau_{1},\tau_{2} )F_{ac}(|{\bf k} - {\bf p}|, \tau_{1},\tau_{3} )\\\nonumber
&&\qquad\qquad +(p-k)^{o}(p-k)^{p}p^{q}p^{r}F_{bc}(|{\bf k}''+{\bf p}|, \tau_{3}, \tau_{2}) F_{ab}(|{\bf k}-{\bf p}|, \tau_{1},\tau_{2} )F_{ac}(p, \tau_{1},\tau_{3})\Big).
\end{eqnarray}
Here the factor for 4 comes from the equivalent diagrams and ${\bf k} = f\hat{\Omega}$ etc. 
\begin{figure}
\centering
\includegraphics[scale = 0.6]{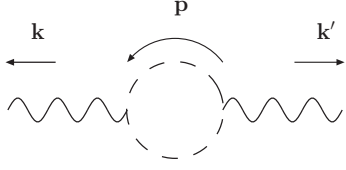}
\caption{The gravitational wave power spectrum generated by a scalar source is given by the 1-loop diagram.} \label{fig:2ptloop}
\end{figure}
The power spectrum can be calculated in the analogous manner. The power spectrum is generated by the diagram in fig. \ref{fig:2ptloop}. We obtain
\begin{eqnarray}\nonumber\label{eqn:2ptlinear}
\langle h_{ij}({\bf k }, \tau)h_{ij}({\bf k}', \tau)\rangle & = & 2\delta({\bf k}+{\bf k}')\mathcal{O}_{kl,mn}({\bf k})\left(\frac{16 \pi G}{k}\right)^{2} \int_{\tau_{i}}^{\tau_{f}}d\tau_{1}\int_{\tau_{i}}^{\tau_{f}}d\tau_{2}a(\tau_{1})\sin(k\tau_{1})a(\tau_{2})\sin(k\tau_{2})\\&&
\times\tau_{1}^{3}\tau_{2}^{3}\int\frac{d^{3}p}{(2\pi)^{3}}p^{k}p^{l}p^{m}p^{n}F_{ab}(p,\tau_{1}, \tau_{2})F_{ab}(|{\bf p}-{\bf k}|,\tau_{1},\tau_{2}).
\end{eqnarray}
This result is consistent with that quoted in \cite{Dufaux:2007pt}.

\bibliography{StochGW1}
\end{document}